\documentstyle[prl,aps,twocolumn,epsfig]{revtex}

\begin{document}
\draft
\title{Backflow in a Fermi Liquid}
\author{W. Zwerger}
\address{Sektion Physik, Universit\"at M\"unchen,
Theresienstra{\ss}e 37, D-80333 M\"unchen, Germany}
\date{\today}
\maketitle

\begin{abstract}
We calculate the backflow current around a fixed impurity in
a Fermi liquid. 
The leading contribution at long distances
is radial and proportional to $1/r^2$. It is caused by the
current induced density modulation first discussed by Landauer.
The familiar $1/r^3$ dipolar backflow obtained 
in linear response by Pines and Nozieres is only the next to
leading term, whose strength is calculated here to all orders in
the scattering. In the charged case the condition of perfect
screening gives rise to a novel sum rule for the phase shifts.
Similar to the behavior in a
classical viscous liquid, the friction force is due only to the
leading contribution in the backflow while the dipolar term does
not contribute. 
\end{abstract}

The calculation of backflow in liquids
is one of the standard problems in hydrodynamics, determining e.g.
the Stokes friction in a classical viscous liquid \cite{Landau}
or the properties of rotons and impurities in superfluid 
Helium $4$ \cite{Feynman}. 
For the case of a Fermi liquid, the backflow around a slowly 
moving massive impurity is discussed in the classic text by
Pines and Nozieres \cite{Pines}. Within linear response they show
that the backflow is proportional to the density response function
and dipolar in character. As pointed out by these authors, the
dipolar form may be derived from a rather simple geometrical
argument: Indeed the backflow current outside the impurity should
have zero divergence being a stationary flow and zero curl because
the perturbation is longitudinal. The only vector function
obeying both conditions, however, is a dipole. For a neutral
Fermi liquid the strength of the dipole is given by the
compressibility times the Fourier transform of the interaction
at zero momentum. In the charged case the dipolar backflow has
a universal amplitude. This is a result of perfect screening
which requires that the backflow identically cancels the
longitudinal part of the impurity current \cite{Pines}.

In this letter we reconsider the backflow problem in a Fermi
liquid, going beyond the linear response treatment. Starting 
with the simple case of a noninteracting Fermi gas, we show that
the leading term at long distances is not the dipolar backflow
but a radial contribution decaying like $1/r^{d-1}$ in $d$
dimensions ($d=2,3$ in the following). It is proportional to
the impurities transport cross section and thus is not contained
in a linear response calculation where the interaction only
appears linearly. The novel term has nonzero curl and is
directly related to the asymmetry in density around localized
scatterers in the presence of a finite transport current, discussed
long ago by Landauer \cite{Landauer}. We also calculate the
familiar dipolar backflow to all orders in the scattering potential
and discuss the modifications of our results for interacting
Fermi liquids. It is found that the condition of
perfect screening entails a sum rule for the scattering phase
shifts which is similar to, but different from the one by
Friedel \cite{Friedel}. Finally we determine the systematic force
exerted on the impurity by the moving fermions.
It is shown that only the leading 
$1/r^{d-1}$ term of the backflow current contributes to the
force, a situation which is completely analogous to
that in a classical viscous liquid.

Let us consider a fixed scattering center at the origin which
is characterized by a spherically symmetric interaction 
potential $V(\vec x)$. In the frame where the impurity is at
rest, the Fermi system is flowing past with asymptotic velocity
$\vec v\neq 0$. The unperturbed current density is therefore
$\vec j(\vec x)|_0=n\vec v$, with $n$ the equilibrium 
number density.
Due to scattering off the impurity, the actual current density
$\vec j(\vec x)$ differs from $n\vec v$ by a backflow current
$\delta\vec j(\vec x)$. To lowest order in $\vec v$ the Fourier
transform $\delta\vec j(\vec q)$ of the backflow is of the form
\begin{equation}
\label{h(q)}
\delta\vec j(\vec q)=h(q)\left[(\hat q\vec v)\hat q-\vec v\right]
\end{equation}
where $\hat q$ is the unit vector in the direction of $\vec q$.
Indeed the vector in Eq.(\ref{h(q)}) is uniquely determined by
the requirement that it is linear in $\vec v$ and the zero
divergence condition $\vec q\cdot\delta\vec j(\vec q)=0$ due to
the stationarity of the flow. For small velocities the 
backflow pattern is thus completely determined by the scalar
function $h(q)$. As pointed out above, a treatment of the
interaction potential $V(\vec x)$ in linear response gives 
rise to a dipolar backflow which is characterized by
$\lim_{q\to 0}h(q)=h_0$. The associated dimensionless constant
$h_0$ is equal to ${\partial n/\partial\mu}\cdot V(q=0)$
in the case of a neutral Fermi liquid \cite{Pines}. Here 
the compressibility $\partial n/\partial\mu$ is just the
$q\to 0$ limit of the general density response function
$\chi(q)$. For an impurity with charge $Z$, the product
$\chi(q)\cdot V(q)$ is replaced by $Z\left[\epsilon^{-1}(q)-1
\right]$ with $\epsilon(q)$ the static dielectric constant
\cite{Pines}. As a result of the perfect screening condition
$\epsilon^{-1}(q\to 0)=0$, this leads to a universal value
$h_0^{c}=-Z$ for the strength of the dipolar backflow in the
charged case.

In order to discuss the generalization of these results beyond
linear response, still keeping the asymptotic velocity $\vec v$
small however, we start by considering a noninteracting Fermi 
gas. In this case the backflow can be calculated analytically
from the single particle eigenstates $\psi_k(\vec x)$, which
are the exact outgoing scattering states in $V(\vec x)$. Indeed
describing the finite asymptotic current $n\vec v$ by a shifted
Fermi distribution $f(\epsilon_{\vec k-m\vec v/\hbar})$ for the
incoming momenta $\vec k$, the total fermionic current density
at zero temperature and to linear order in $\vec v$ is given by   
\begin{equation}
\label{j(q)}
\vec j(\vec x)=\frac{k_F^{d-1}}{(2\pi)^d}\int d\Omega_k\,
\hat k\cdot\vec v\;\hbox{Im}\left[\,\psi_k^*(\vec x)\nabla_x
\psi_k(\vec x)\right] |_{k=k_F}\; .
\end{equation}
Here $d\Omega_k$ denotes an integration over the directions of the
unit vector $\hat k$, while the magnitude $k=|\vec k|$ is
fixed at the Fermi wave vector $k_F$. Thus at $T=0$ and to linear
order in $\vec v$ the backflow is completely determined by
the exact scattering states right at the Fermi surface. Clearly
the behavior of $\vec j(\vec x)$ at arbitrary distances depends on
the details of $\psi_k(\vec x)$. For large distances, however,
it is sufficient to know the asymptotic form of the scattering
states. In order to obtain the first two leading contributions
to $\vec j(\vec x)$ it is necessary to expand
\begin{equation}
\label{psi3}
\psi_k(\vec x)\;\to\; e^{i\vec k\vec x}+f\cdot
\frac{e^{ikr}}{r}+f_2\cdot\frac{e^{ikr}}{r^2}+\ldots
\end{equation}
to order $1/r^2$ in three dimensions. Here $f$ is the standard
scattering amplitude while the coefficient of the $1/r^2$
contribution is given by
\begin{equation}
\label{f2ind=3}
f_2=\frac{i}{2k^2}\sum_{l=0}^{\infty}(2l+1)l(l+1)e^{i\delta_l}
\sin{\delta_l}\, P_l(\hat k\cdot\hat x)
\end{equation}
with phase shifts $\delta_l$ and the usual Legendre polynomials
$P_l$. This result is obtained by a straightforward asymptotic 
expansion of
the free particle solutions with given angular momentum $l$. In
two dimensions the corresponding result turns out to be
\begin{equation}
\label{psi2}
\psi_k(\vec x)\;\to\; e^{i\vec k\vec x}+f\cdot
\frac{e^{ikr}}{r^{1/2}}+f_2\cdot\frac{e^{ikr}}{r^{3/2}}+\ldots
\end{equation}
with amplitudes $f$ and $f_{2}$ which are not given here explicitely.
It is now straightforward to insert the asymptotic behavior of
the scattering states into our expression (2) for the current.
Apart from the trivial term $\vec k$ which accounts for the
background current density $n\vec v$,
the leading contributions to Im$\left[\psi^*\nabla\psi\right]
|_{k=k_F}$ obviously arise from the square $k_F\hat x|f|^2/
r^{d-1}$ of the outgoing wave and the two interference terms 
linear in $f$. Now $\exp{i(kr-\vec k\vec x)}$ is asymptotically
proportional to a $\delta$-function $\delta(\Omega_k-\Omega_x)$
which singles out the forward direction $\hat k=\hat x$. Using
the optical theorem it is straightforward to show that the
leading term to the backflow is given by \cite{Bonig}
\begin{equation}
\label{leading j}
\delta\vec j(\vec x)\to\; -\frac{k_F^d}{(2\pi)^d}
\frac{\sigma_{tr}}{r^{d-1}}(\hat x\vec v)\hat x\; +O(r^{-d})
\end{equation}
with $\sigma_{tr}=\int d\Omega_k(1-\hat k\hat x)|f|^2$ the
standard transport cross section. Obviously the contribution
(6) is a purely radial current which vanishes in the direction
perpendicular to $\vec v$ (see Fig.1). It has vanishing 
divergence as it should, but finite curl. 
\begin{figure}
\centering\epsfig{file=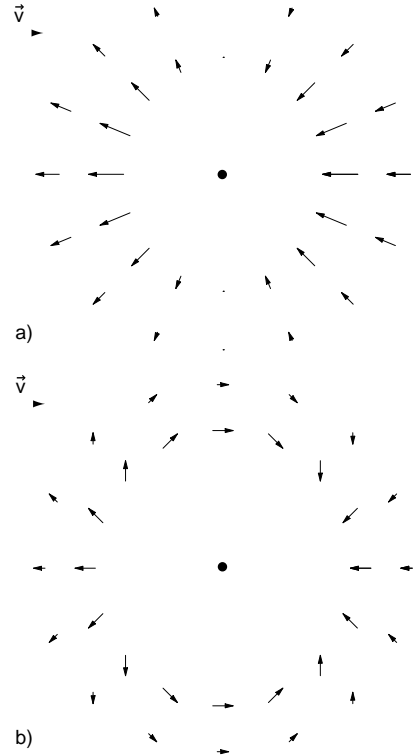,scale=0.53}
\caption{\label{fig1}a) Radial backflow current 
$-(\hat{x}\vec{v})\hat{x}/r$ in $d=2$. The impurity sits at the center 
with incoming current from the left.
b) Dipolar backflow $\left(\vec{v}-2(\hat{x}\vec{v})\hat{x}\right)/r^{2}$
in $d=2$ for a repulsive impurity at the center. The direction of the 
flow is reversed in the attractive case, in contrast to a).}
\end{figure}
In order to 
understand its physical origin we consider the current induced
part $\delta n(\vec x)$ of the density modulation which is 
caused by the scattering off the impurity. As predicted by
Landauer \cite{Landauer} this modulation asymptotically has the form
$\delta n(\vec x)\sim -(\hat x\vec v)/r^{d-1}$ of a dipole
potential. Comparing the exact expression obtained for
$\delta n(\vec x)$ in a scattering theory calculation
\cite{Zwerger} with our result (6), it turns out that at $T=0$
and to linear order in $\vec v$ the asymptotic
backflow current is simply given by
\begin{equation}
\label{jdelta n}
\delta\vec j(\vec x)=v_F\delta n(\vec x)\cdot\hat x\; .
\end{equation}  
The leading term in the backflow is thus directly proportional
to the current induced density change $\delta n(\vec x)$
which is positive in front and negative behind the scatterer,
in agreement with the intuitive picture developed by Landauer
\cite{Landauer}. As a result, the sign of this contribution to
the backflow remains unchanged upon going from a repulsive to
an attractive potential $V(\vec x)$. This is in contrast to the
dipolar contribution  
\begin{equation}
\label{dipolar j}
\delta\vec j(\vec x)|_{dip}=-\frac{h_0}{2\pi(d-1)}\,
\frac{d(\hat x\vec v)\hat x-\vec v}{r^d}
\end{equation}
which is only the next to leading term in an asymptotic expansion
of $\vec j(\vec x)$. Using (3) and (5) a straightforward but
rather tedious calculation indeed gives a contribution to
$\delta\vec j(\vec x)$ of the form (8) with strength 
\begin{equation}
\label{h0}
h_0=\frac{\partial n}{\partial\mu}\cdot\hbox{Re}\, V_0\, -\,
\frac{2}{\pi}\sum_{l=0}^{\infty}c_l\sin{\delta_l}
\sin{\delta_{l+1}}\sin{(\delta_l-\delta_{l+1})}\; .
\end{equation}
Here $c_l=2l+1$ or $(l+1)^2$ in $d=2$ or $d=3$ respectively, 
while
\begin{equation}
\label{V0}
V_0=\int d^dx\, e^{-i\vec k\vec x}V(\vec x)\psi_k(\vec x)
\end{equation}
is essentially the exact forward scattering amplitude. Since
$V_0$ reduces to $V(q=0)$ in Born approximation, the first term
in (9) is the obvious generalization of the linear response
result of Pines and Nozieres to arbitrary order in the scattering
potential. It is convenient to express also this contribution in
terms of the phase shifts $\delta_l$ via
\begin{equation}
\label{h01}
h_0^{(1)}=-\frac{1}{2\pi}\sum_{l=0}^{\infty}a_l\,
\sin{2\delta_l}
\end{equation}
where $a_l=2-\delta_{l,0}$ or $2l+1$ in $d=2,3$.
In addition to $h_0^{(1)}$ there is a second contribution
which is at least of order $V^3$. It arises from the
interference term $k_F\hat x/r^d\,\cdot 2\hbox{Re}\, f^*f_2$
between the first and next to leading contribution to the 
outgoing wave. The additional term is odd under $\delta_l\to
-\delta_l$ as is the first one, but vanishes in the case of
s-wave scattering only. 

Including both the radial and
dipolar contributions to the
backflow current, the scalar function $h(q)$ defined in (1) has
the general form
\begin{equation}
\label{limh(q)}
\lim_{q\to 0}\, h(q)=\frac{h_{-1}}{q}+h_0+\ldots
\end{equation}
with $h_{-1}=(2,3\pi/4)n\sigma_{tr}$ in $d=(2,3)$. Since
\begin{equation}
\label{sigmatr}
\sigma_{tr}=\frac{4}{k_F^{d-1}}\sum_{l=0}^{\infty}b_l\,
\sin^2{(\delta_l-\delta_{l+1})}
\end{equation}
with $b_l=1$ or $\pi(l+1)$ in $d=2,3$, both leading coefficients
$h_{-1}$ and $h_0$ can be expressed completely
in terms of the density $n\sim k_F^d$ and 
the scattering phase shifts $\delta_l$. In the particular case
of $d=3$ and s-wave scattering with scattering length $a$,
we have $h_{-1}=(k_Fa)^2k_F/2$ and $h_0=k_Fa/\pi$. With $k_F$ as 
the typical scale for $q$, this shows that the strength of the
leading radial term is then a factor $k_Fa\ll 1$ smaller than the
dipolar contribution. Nevertheless at long distances it is 
always the radial term which dominates.

In a Fermi liquid the interacting state develops adiabatically from
the noninteracting one. The resulting quasiparticle states in
a local and centrally symmetric potential are characterized by phase 
shifts $\delta_{l}$. In general these are functionals of both the 
energy and the quasiparticle distribution $n_{k}(\vec x)$
\cite{Nozieres}.
By Galileian invariance the asymptotic distribution
is again a Fermi sphere shifted by $\delta\vec k=m\vec v/\hbar$,
with $m$ the bare mass. At $T=0$ the energy is fixed at $\epsilon_{F}$
and there are no collisions other than with the impurity. Moreover
since the deviation of $n_{k}$ from equilibrium is already linear in
$\vec v$, we may neglect the dependence of $\delta_{l}$ both on
energy and on $n_{k}$. The resulting values of $\delta_{l}$ then
define an effective force $\vec F_{k}$ on the quasiparticles which
appears in the corresponding transport equation \cite{Pines}. In
a fully quantum mechanical treatment of the Wigner function 
$n_{k}(\vec x)$ the associated local particle current must then
be equal to $\vec j_{k}(\vec x)=\frac{\hbar}{m}\hbox{Im}\,
\psi_{k}^{*}\nabla_{x}\psi_{k}$ where $\psi_{k}$ are the exact
scattering states in an effective potential with phase shifts
$\delta_{l}$. As was shown above, $h_{-1}$ 
and $h_0$ can be expressed completely in terms of $k_F$
and the scattering phase shifts $\delta_l$.  The
generalization of our results to the
interacting case is therefore rather obvious. Indeed since 
$k_F$ is unchanged  
one only needs to replace the phase shifts by those for
quasiparticles. The general form of the backflow as determined
by (1) and (12) thus applies also in the interacting case, however
with renormalized parameters $h_{-1}$ and $h_0$. For a charged
impurity in an electron liquid, the perfect screening
condition must hold to all orders in $V$. As we have seen, this
implies a universal dipolar backflow characterized by 
$h_0^{c}=-Z$ for an impurity with charge $Z$.
Since $h_0$ is completely determined by the $\delta_l$ via
(9) and (11), perfect screening 
gives a nontrivial condition on the scattering phase shifts at
a charged impurity. In the limit $\delta_{l}\ll 1$
it reduces to the well known Friedel sum
rule \cite{Friedel} which fixes the number of bound states.
The novel sum rule shows that even for $Z=1$
no purely s-wave scattering potential
can account for the backflow in the charged case. Regarding
the dominant radial contribution, the transport cross section
appearing in the coefficient $h_{-1}$ has 
to be replaced by its value for the screened potential
$V(q)/\epsilon(q)$. In contrast to $h_0$ the strength of the
radial backflow is therefore not universal.

Finally we calculate the systematic force $\vec F$ due to the transfer
of momentum between liquid and scatterer. In the context
of electromigration theory this is known as the wind force
\cite{Verbruggen}. Taking the gradient of the interaction energy
with respect to the impurity position, it is straightforward to
see that
\begin{equation}
\label{F1}
\vec F=-\int d^{d}x\, n(\vec x)\nabla_{x}V\; .
\end{equation}
Clearly at zero current $\vec v=0$ this force vanishes although the 
fermion density is not uniform even in this case. Therefore only the
current induced density change $\delta n(\vec x)$ contributes to
$\vec F$. For simplicity we consider again a Fermi gas at $T=0$ with  
scattering states $|\vec k+>$. 
To lowest order in $\vec v$ the current induced density then has
the asymptotic behavior ($d=2,3$)
\begin{equation}
\label{dn}
\delta n(\vec x)=-\frac{1}{2\pi 
v_{F}}\left[\frac{(1,2/\pi)h_{-1}}{r^{d-1}}+\frac{h_{0}}{r^{d}}+\ldots
\right]\,\hat x\vec v
\end{equation}
which shows that the two leading terms in $v_{F}\delta n(\vec x)$ are 
identical with the radial component $\hat x\cdot\delta\vec j(\vec x)$
of the backflow. The associated total force can be written as
\begin{equation}
\label{F}
\vec F=\frac{k_F^{d-1}}{(2\pi)^d}\int d\Omega_k\,
\hat k\cdot\vec v\;\frac{m}{\hbar}<\vec k+|-\nabla_xV|
\vec k+>|_{k=k_F}
\end{equation}
similar to (2). Now the relevant matrix element of $\nabla_{x} V$ between
the exact scattering states is equal to
$2\epsilon_{F}\sigma_{tr}(k_F)\cdot\hat k$. Thus (16)
immediately gives a conventional friction force $\vec F=
-\eta_F\vec v$ with $\eta_F=\hbar k_Fn\sigma_{tr}$ \cite{Bonig}.
The fermionic friction coefficient $\eta_F$ is  
proportional to the transport cross section which appears in the
radial contribution $h_{-1}$ to the backflow. It is this term which
determines the single impurity contribution to the residual
resistivity \cite{Landauer,Zwerger}. This is a simple example of
the so called Das-Peierls theorem \cite{Verbruggen,Das} in 
electromigration, which states that the total force on the impurity
is proportional to the additional resistivity it causes. The fact that
the dipolar contribution $h_{0}$ to the backflow does not contribute
to the friction force can be understood most easily by considering
the linear response regime. Indeed to linear order in $V$ the
response at low velocities is purely reactive \cite{Pines},
while a finite resistivity can only appear at order $V^{2}$.
More generally, the coefficient $h_{0}$ is odd in $\delta_{l}$,
while the force must be an even function of the phase shifts.
This situation is in fact very similar
to the case of a classical, incompressible and viscous liquid.
Calculating the backflow current around a sphere of radius $R$
with boundary condition $\vec v=0$ at the surface,
one finds \cite{Landau} that $\delta\vec j(\vec x)$ has
a contribution proportional to $1/r$ and a dipolar one. The
associated function $h(q)$ as defined in (1) is thus of the form
\begin{equation}
\label{hcl}
\lim_{q\to 0}\, h_{cl}(q)=\frac{h_{-2}}{q^2}+h_0+\ldots\; .
\end{equation}  
The coefficient of the $1/r$ contribution is $h_{-2}=6\pi Rn$
while the strength of the dipolar backflow is negative and given
by $h_0=-\pi R^3n$ (the corresponding problem in two 
dimensions has no solution which is known as the Stokes
paradox). Calculating the associated friction
$\vec F=-\eta_S\vec v$ in a fluid with kinematic viscosity $\nu$
it turns out \cite{Landau} that only the leading term $h_{-2}$
contributes to
$\eta_S=6\pi Rn\cdot m\nu$
while the dipolar backflow again drops out.
Comparing the Stokes result with that for a Fermi liquid, we see
that the fermionic friction coefficient for a scattering potential with
characteristic range $R$ such that $\sigma_{tr}=\pi R^2$
is equal to that of a classical liquid
with finite kinematic
viscosity $\nu_F=v_FR/6$. With
typical values $R=2\,\AA$ and $v_F=1.5\cdot 10^8\,\hbox{cm/sec}$ 
for electrons in metals, we obtain $\nu_F=0.5\,\hbox{cm}^2/
\hbox{sec}$ which
is about fifty times the viscosity of water. From this
point of view therefore, electrons in metals behave like a 
rather viscous liquid indeed !

In summary we have calculated the backflow around a fixed
impurity in a Fermi liquid at low velocities and zero temperature.
The dominant contribution is radial
and decays like $1/r^{d-1}$. It
is directly proportional to the current induced density 
modulation first discussed by Landauer and is responsible for
the frictional force i.e. the finite resistivity.
The subleading dipolar contribution of Pines and Nozieres has been
evaluated to arbitrary orders in the scattering. 
It has been shown that the condition of perfect
screening of a charged impurity gives rise to a novel sum rule
for the corresponding phase shifts. We have also evaluated the
total force on the impurity. In agreement with the Das-Peierls
theorem it is proportional to the scatterers contribution to
the resistivity. For our simple situation this is just a
consequence of Newtons third law. The fact that the dipolar 
term in the backflow gives no contribution to the force 
shows that the so called direct force in
electromigration theory \cite{Verbruggen,Bosvieux} vanishes in  
the present case. This is a consequence of the way we 
have set up the problem: Instead of calculating a current as 
the response to an external field we have specified the
incoming current which gives rise to a certain backflow or
potential distribution \cite{Landauer}. It is an interesting 
future problem to develop a theory of electromigration from
this point of view, including the lattice, background scattering etc. 
\acknowledgments
Constructive comments by H. Wagner are gratefully acknowledged.
This work has been supported by the SFB 348.

\end{document}